\documentclass[pra, 10pt, twocolumn, aps, superscriptaddress, groupaddress]{revtex4}
\usepackage[T1]{fontenc}
\usepackage{bigints}
\usepackage{amssymb}
\usepackage{graphicx}
\usepackage{amsmath}
\usepackage{dcolumn}
\usepackage{multirow}
\usepackage{hyperref}
\usepackage[normalem]{ulem}
\usepackage{physics}
\usepackage{bm}
\usepackage{comment}
\usepackage{float}
\usepackage{xcolor}
\usepackage[caption=false]{subfig}
\usepackage{siunitx}
\usepackage{dsfont}
\usepackage{bbold}
\usepackage[toc]{appendix}
\usepackage{relsize}
\usepackage{leftidx}
\usepackage{bbm}
\usepackage{stackengine}
\usepackage{mathtools}

\newcommand{\be}{\begin{equation}}
\newcommand{\ee}{\end{equation}}
\newcommand{\ba}{\begin{array}}
\newcommand{\ea}{\end{array}}
\newcommand{\bqa}{\begin{eqnarray}}
\newcommand{\eqa}{\end{eqnarray}}

\definecolor{lapislazuli}{rgb}{0.15, 0.38, 0.61}
\definecolor{YKblue}{rgb}{0.0, 0.18, 0.65}
\definecolor{carmine}{rgb}{0.81, 0.09, 0.13}
\definecolor{lavender}{rgb}{0.84, 0.79, 0.87}

\hypersetup{
	colorlinks=true,
	linkcolor=YKblue,
	linktoc=page,
	citecolor=blue,
	urlcolor=YKblue
}

\renewcommand{\vec}[1]{\bm{#1}}
\renewcommand{\grad}{\bm{\nabla}}
 
\DeclareMathOperator{\PolyLog}{Li}
\newcommand{\Li}[1]{\PolyLog_{#1}}
\DeclareMathOperator{\PlasmaZ}{Z}
\newcommand{\plasma}[2]{\PlasmaZ^{\rm{#1}}_{#2}}

\begin{document}

\title{Anisotropy-Induced Magnetic Field Generation in Bidimensional Materials}
\date{26th June 2025}
\author{Diogo Simões}
\affiliation{GoLP/Instituto de Plasmas e Fusão Nuclear, Instituto Superior Técnico, Lisboa, Portugal}
\email{diogo.s.simoes@tecnico.ulisboa.pt}
\author{Hugo Terças}
\affiliation{Instituto Superior de Engenharia de Lisboa, Instituto Politécnico de Lisboa, Lisboa, Portugal}
\affiliation{GoLP/Instituto de Plasmas e Fusão Nuclear, Instituto Superior Técnico, Lisboa, Portugal}
\author{Jorge Ferreira}
\affiliation{Instituto de Plasmas e Fusão Nuclear, Instituto Superior Técnico, Lisboa, Portugal}


\begin{abstract}

    We investigate an electromagnetic instability in two-dimensional materials arising from an anisotropy of the Fermi surface, utilizing a kinetic model accounting for the effects of the values of temperature, chemical potential and anisotropy ratio, as well as considering both linear and quadratic low-energy band structures. The wavenumber-dependent growth-rate of these modes is derived in the linear regime, and their confinement, contrasting with stable electromagnetic waves in these systems, is described. The generation of structured out-of-plane magnetic fields, as well as their behaviour in saturation, is shown using fully kinetic and non-linear simulations.\\

\end{abstract}

\maketitle


\section{Introduction}
\label{sec:intro}

In recent years, bidimensional materials have proven to be a very active area of study, in both theoretical physics as well as experimental physics and engineering, with numerous applications, due to their capability to realize novel optical, electronic and mechanical properties \cite{graph_props}. These materials are characterised by the confinement of the wavefunctions of the charge carriers to the plane. Beyond this, the high concentration of conductors, as well as the modification of the Coloumb interactions between them due to this geometry, drives a plethora of interesting phenomena.

In the hydrodynamic spirit, graphene field-effect transistors (GFETs) have been proposed as platforms for excitation of plasmonic waves \cite{Aizin1}, which can be used to generate and detect THz waves \cite{cosme_combs,cosme_feedback}, constituting a candidate for closing the THz gap, with important applications for sensing \cite{sensing}, medical imaging \cite{medical}, among others. Furthermore, non-linear \cite{cosme_nonlinear} and chiral waves \cite{cosme_chiral} have been theorised, with potential applications in communications \cite{comms}, in setups involving magnetic domain boundaries.

Monolayer and bilayer graphene are representative examples of linear-band and quadratic-band two-dimensional materials, respectively, however, the exact nature of the bands of a material is here ignored. We assume only that the behaviour around the Fermi energy is known, and we focus on materials where this behaviour is linear or quadratic, as these two cases cover a large number of two-dimensional conductive materials, and the exact coefficients of the bands $v_F$ or $m$ will be left as variables in the analytical work, and adimentionalized around in the simulations, so as to keep the analysis as general as possible.

The field of plasmonics in these systems has focused on transverse magnetic (TM) waves, as transverse electric (TE) waves have been shown to not be supported, with the exception of narrow frequency bands due to interband transitions \cite{Peres4}.
In spite of this, this paper focuses on a TE instability arising in bidimensional materials with anisotropic band structure, which is supported due to its non-oscillating nature. This instability is shown to generate out-of-plane magnetic fields, which may be tunable by changing parameters such as the chemical potential with a gate.

While the equilibrium, ballistic \cite{ballistic} and hydrodynamic \cite{hydrodynamic} regimes of these systems are well studied, a quantum-kinetic model is here presented and used, so as to allow an analysis beyond the linear regime, where the assumption of local equilibrium cannot be justified.


\section{Kinetic Model}
\label{sec:backg}

With the objective of modelling the carrier dynamics in these systems without resorting to the local equilibrium assumption inherent in a hydrodynamic model, a quantum kinetic description is used, based on the Wigner function,
\begin{subequations}
    \begin{equation}
        W(\vec{r},\vec{p}) = \frac{1}{(2\pi\hbar)^d}\int \bra{\vec{r}-\vec{s}/2}\hat \rho\ket{\vec{r}+\vec{s}/2} e^{i \vec{p}\cdot\vec{s}/\hbar} d\vec{s},
        \label{eq:wigner_xdef}
    \end{equation}
\end{subequations}
with $\hat \rho$ the density operator.
Starting from this operator's definition and time-evolution equation, and using the above defined Wigner transform, the time-evolution of this quasiprobability distribution can be written as \cite{moyal1949},
\begin{equation}
    \begin{split}
        &\frac{\partial W}{\partial t} = \{\{H,W\}\}\\
        &\coloneqq \frac{2}{\hbar} W(\vec{r},\vec{p},t) \sin\left(\hbar\frac{\overleftarrow{\grad}_p\cdot\overrightarrow{\grad}_r - \overleftarrow{\grad}_r\cdot\overrightarrow{\grad}_p}{2}\right) H(\vec{r},\vec{p}).
        \label{eq:moyal}
    \end{split}
\end{equation}
The classical limit of this equation is the Liouville equation, and, neglecting correlations, the single-particle Vlasov equation is obtained after truncation of the BBGKY hierarchy,
\begin{equation}
    \partial_t f(\vec{r},\vec{p},t) + \Vec{v}\cdot\Vec{\nabla_r} f(\vec{r},\vec{p},t) + \Vec{F}\cdot\Vec{\nabla_p} f(\vec{r},\vec{p},t)=0,
    \label{eq:vlasov}
\end{equation}
with $\vec{v} = \grad_p E(\vec{p})$ the microscopic particle velocity, dependent on the band defined by $E(\vec{p})$, and $\vec{F}$ the Lorentz force. 


\section{Numerical Implementation}
\label{sec:imple}

Having established the kinetic equation which rules the time-evolution of the system, it is of interest to create non-linear numerical simulations to go beyond analytically achievable results.
As such, a discretization of this partial differential equation (PDE), along with an algorithm to solve it numerically is developed. These efforts resulted in the creation of a simulation code entitled {\bf Quetzal} (QUantum Electrodynamic bolTZmann ALgorithm).

The algorithm consists of a finite-volume scheme, acting on a phase-space cartesian grid. The fluxes are estimated with a monotone upstream scheme, from piecewise linear reconstruction of cell values coupled to a monotonized-central flux limiter, yielding a second-order accurate flux. The time evolution is carried out by a second-order accurate Runge-Kutta total variation diminishing (TVD) method.

The fields, on the other hand, are evolved in time via the same TVD Runge-Kutta method, and their time derivatives are estimated via a spectral method, calculating the necessary convolutions of the current density with a kernel, which will be derived below, via the fast Fourier transform (FFT) algorithm.


\section{Modified Maxwell system}
\label{sec:resul}

The electromagnetic fields are governed by the time-dependent Maxwell system of equations, however, the 2D geometry of the system, with charge confined to the material plane, changes their evolution.
To calculate the governing equations under this geometry, the Fourier transform of the fields along the planar axes ($\hat x$ and $\hat y$) is taken, and the system of equations is solved for the waves in the dielectric medium above and below the material's plane, which take the form of incoming and outgoing waves of either polarization, yielding a total of 4 unknown constants for either region.

Then, the unknown constants are set by imposing boundary conditions at the plane ($z=0$),
\begin{subequations}
\begin{equation}
    E_{x,y}(z=0^+) = E_{x,y}(z=0^-),
\end{equation}
\begin{equation}
\begin{split}
    B_x(z=0^+)-B_x(z=0^-) = \mu J_y(\vec{q}),\\
    B_y(z=0^+)-B_y(z=0^-) = -\mu J_x(\vec{q}),\\
\end{split}
\end{equation}
\end{subequations}
corresponding to the continuity conditions of the electric field and and discontinuity condition for the magnetic field due to the current density $\vec{J}_{\rm 3D}(x,y,z) = \vec{J}(x,y) \delta(z)$, as well as the assumption that no incoming waves from $z\xrightarrow{}\pm\infty$ exist.
This process results in the system of equations given by
\begin{subequations}  \label{eq:2DMaxwell}
    \begin{equation}
       \partial_t E_x = -i q_y B_z - \kappa_z \frac{J_x}{2},
       \label{eq:2DMaxwellEx}
    \end{equation}
    \begin{equation}
      \partial_t E_y = i q_x B_z - \kappa_z \frac{J_y}{2},
      \label{eq:2DMaxwellEy}
    \end{equation}
    \begin{equation}
        \partial_t B_z = -i q_y E_x + i q_x E_y,
        \label{eq:2DMaxwellBz}
    \end{equation}
\end{subequations}
with $\kappa_z = \sqrt{q^2 -\omega^2/c^2}$ being the decay rate of the fields along $\hat z$. 

The general process to calculate dispersion relations or growth-rates involves taking the Fourier transform of the Vlasov equation (eq. \ref{eq:vlasov}), and linearizing around the appropriate equilibrium $f_0$. This process results in the first-order perturbation
\begin{equation}
    \Tilde{f_1} = i \frac{\Vec{\Tilde{F_1}}}{\omega - \vec{v}\cdot\vec{q}}\cdot\Vec{\nabla_p} f_0(\vec{p}).
\end{equation}
We assume here that no permanent zero-order current is driven in the material, such that the current is given by an integral over this perturbation to the distribution function.

With this result, we search electromagnetic modes polarized with the magnetic field along $\hat{z}$ and in-plane electric field, confined to the material, by inserting this result into the system of equations \ref{eq:2DMaxwell}. This leads to the secular equation
\begin{equation}
    \begin{split}
        1 = \frac{1}{\kappa_z} \frac{\omega e^2}{2 c^2 \epsilon}\Bigg[\frac{q}{\omega}\int \frac{v_y^2}{\omega-v_x q}\frac{\partial f_0}{\partial p_x} d^2\vec{p}\\
        +\int v_y \frac{1 - v_x \frac{q}{\omega}}{\omega-v_x q}\frac{\partial f_0}{\partial p_y} d^2\vec{p}\Bigg].
    \end{split}
    \label{eq:EMsecular}
\end{equation}
With an equilibrium Fermi-Dirac distribution and quadratic bands, the dispersion relation obtained is
\begin{equation}
    \omega^2 = c^2 q^2 - g_e \kappa_z \zeta\plasma{2DFD}{1}(\zeta),
\end{equation}
with $g_e= \frac{e^2 n_e}{2 \varepsilon m}$, $n_e$ the electron density, $\zeta=\omega/(q v_{\rm th.})$ and $\plasma{2DFD}{1}(\zeta)$ the two-dimensional and Fermi statistics analog of the classic plasma dispersion function. This is given by the integral
\begin{equation}
   \plasma{FD}{n}(\zeta) =  - \frac{e^{-\alpha}}{\sqrt{\pi}}\int \frac{1}{s-\zeta}\Li{n-\frac{1}{2}}(-e^{-s^2+\alpha})ds,
   \label{eq:Z_FD}
\end{equation}
which, via appropriate power expansions can be written as the series
\begin{subequations}
    \begin{equation}
    \plasma{FD}{n}(\zeta) = \frac{e^{-\alpha}}{\sqrt{\pi} \zeta} \sum_{l=0}^\infty \frac{1}{\zeta^{2l}} \Gamma(l+\frac{1}{2}) \Li{l+n}(-e^{\alpha}) + \rm{P.C. },
    \label{eq:plasmaZFD_large}
\end{equation}
\begin{equation}
    \plasma{FD}{n}=\sum_{k=1}^{\infty}\frac{(-1)^{k-1} \sqrt{\pi} e^{-\alpha}}{\Gamma(k+\frac{1}{2})} \Li{n-k}(-e^{\alpha}) \zeta^{2k-1} + \rm{P.C. },
    \label{eq:plasmaFDZ_small}
\end{equation}
\end{subequations}
where P.C. is the imaginary part given by the pole contribution to the integral, useful for the large- and short-wavelength approximations via truncation, respectively.

The plasma dispersion function has a negative real part, and considering that $\kappa_z = \sqrt{q^2-\omega^2/c^2}$ is the decay rate of the electromagnetic fields along the direction perpendicular to the material, we can see that we obtain an imaginary value for this quantity. As such, under this model, waves of this nature are not confined in bidimensional materials, and  propagate in the $\hat{z}$ direction.

\section{Anisotropy Instability}

In this section,we aim to exemplify an electromagnetic instability, namely the two-dimensional version of the well-known Weibel instability \cite{Weibel,Fried}, considering linear and quadratic band structures for the charge carriers.

This instability arises due to an anisotropy in momentum-space, frequently realized in classical plasmas as an anisotropy in the thermal velocity of particles along each cartesian direction, introducing an axis-dependent temperature to the equilibrium distribution function.
In a solid-state plasma however, it might prove difficult to realize such an anisotropy in the charge carriers. Furthermore, due to their highly dense nature, it is unlikely that such an anisotropy can be maintained, due to the presence of collisions and their relaxation of the distribution function towards equilibrium. Finally, it is important to note that in these systems charge carriers are typically very cold, with a ratio $\mu/T\gg 1$, making a temperature anisotropy have little effect on the overall velocity distribution of these particles. As such, it is much more feasible to realize this effect with an anisotropy in the chemical potential, i.e., an anisotropic Fermi surface, as is featured in materials such as trilayer graphene, which has a triangular Fermi surface \cite{trilayer}, among others \cite{anisotropic1,anisotropic2,anisotropic3}, due to their angle-dependent band structure. 

We consider a simple model of the equilibrium distribution function, namely one with an elliptical Fermi surface,
\begin{equation}
    f_0 = \left[1 + \exp\left(\frac{p_x^2}{2 m T} + \frac{\mu_x}{\mu_y}\frac{p_y^2}{2 m T}-\frac{\mu_x}{T}\right)\right]^{-1},
\end{equation}
for quadratic bands and
\begin{equation}
    f_0 = \left[1+\exp\left(\frac{v_F}{T}\sqrt{p_x^2+\frac{\mu_x^2}{\mu_y^2}p_y^2}-\frac{\mu_x}{\mu_y}\right)\right]^{-1}.
\end{equation}
for linear ones.

\begin{figure}
    \centering
    \includegraphics[width=.5\columnwidth]{"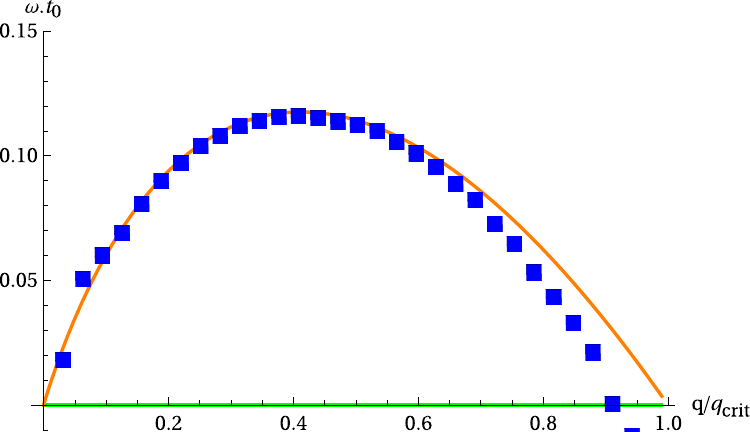"}%
    \includegraphics[width=.5\columnwidth]{"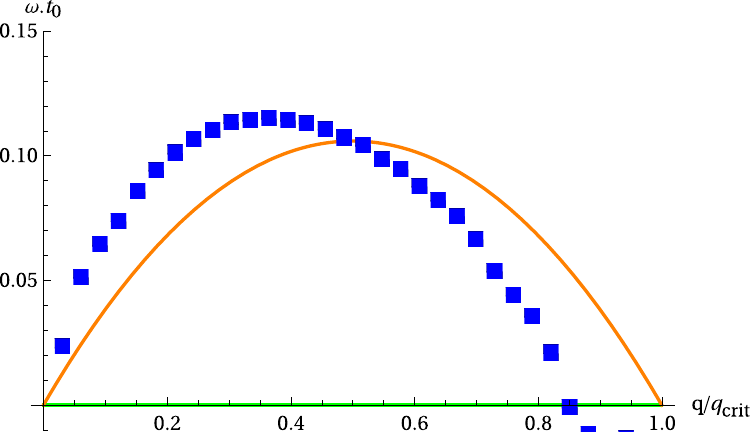"}
    \caption{Growth-rate of the two-dimensional Weibel instability for a quadratic (left) and linear (right) band material as a function of wavenumber $q$ for a chemical potential anisotropies of $\mu_y/\mu_x = 3$, $5$, respectively, and a Fermi-Dirac distribution with shape parameter $\alpha=\mu_x/T=10$. Simulation results in blue.}
    \label{fig:weibel1}
\end{figure}

Firstly considering the quadratic band case, we use eq. \ref{eq:EMsecular}, as this result is general when it comes to distribution function, making only the assumption that it is symmetric along $\hat{y}$.
This leads to the implicit dispersion relation
\begin{equation}
    \omega^2 = c^2q^2 + g_e \kappa_z \left[\frac{n_e}{n_0} - \frac{\mu_y}{\mu_x} \left(\frac{n_e}{n_0}+\zeta \plasma{FD}{1}(\zeta)\right)\right],
    \label{eq:weibel_disp}
\end{equation}
where $\omega$ appears on the RHS both as a part of the decay coefficient $\kappa_z=\sqrt{q^2-\omega^2/c^2}$ and $\zeta=\frac{\omega}{q v_{\rm{th.}}}$.
As we expect a relatively low growth-rate for a finite wavenumber, an assumption to be confirmed with numerical solutions, the plasma dispersion function is approximated with its series around $\zeta=0$, truncated after 5 terms for numerical purposes.
Solving this equation numerically yields the dispersion relation pictured in the left panel of Fig. \ref{fig:weibel1}.

As can be readily seen, the real part of the frequency of this mode is zero, and as such, it represents a purely-growing stationary wave. By plugging in the value $\omega=0$ into eq. \ref{eq:weibel_disp}, the values of wavenumber resulting in a growing mode can be found, yielding the region between zero and $q_{\rm{crit}} = \frac{e^2 n_e}{2 m \varepsilon c^2}(\frac{\mu_y}{\mu_x}-1)$, with a maximum of the growth-rate approximately near the middle of this region. Simulations were performed with one position axis and two momentum axes, due to the transverse nature of the wave and therefore of the movement of the particles to the wavevector, in the linear stage of the instability, exhibiting good agreement with the theoretical results, with a deviation at higher wavenumbers, where the approximation $\zeta\ll1$ fairs worst.

Considering now linear bands, it is useful to change to a system of polar coordinates, taking advantage of the fact that in this a system the particle velocity is independent from the absolute value of the momentum. In this case, we choose a system of elliptical coordinates, so as to both take advantage of the dependence of the velocity only on an angular coordinate, and to make the distribution function only dependent on the radial coordinate, defined as
\begin{equation}
    \begin{split}
        \sigma = \frac{v_F}{T} \sqrt{p_x^2 + \frac{\mu_x^2}{\mu_y^2}p_y^2}\\
        \theta = \arctan\left(\frac{\mu_x}{\mu_y}\frac{p_y}{p_x}\right).
    \end{split}
\end{equation}
A series expansion is used on the integrand of the first integral of \ref{eq:EMsecular}, around the value $\omega/(q v_F) = 0$, invoking the short-wavelength limit as in the quadratic-band case. Keeping the first two terms of this expansion the dispersion relation takes the approximate form
\begin{equation}
    \begin{split}
        \omega^2 = c^2 q^2 + \frac{e^2 n_e v_F^2}{2 \varepsilon T} \kappa_z \frac{\Li{1}(-e^\alpha)}{\Li{2}(-e^\alpha)}\Big[2 E_0(\gamma)\\
        - \gamma^2\left(2 E_0(\gamma)\pm i  \frac{\omega}{q v_F}\right)\Big],
    \end{split}
\end{equation}
with $\Li{n}$ the polylogarithm function and $\gamma = \mu_y/\mu_x$ the anisotropy ratio.
This implicit relation for $\omega$ is similar to the quadratic-band case, with a more complicated dependency on the anisotropy parameter $\gamma$ encoded in the elliptic function,
\begin{equation}
        \mathrm{E}_0(\gamma) = \frac{\mathrm{E}\left(\arccsc(\frac{\gamma}{\sqrt{\gamma^2-1}}); \frac{\gamma^2}{\gamma^2-1}\right)}{\pi \sqrt{\gamma^2-1}},
\end{equation}
where $\mathrm{E}(\phi | k^2) = \int_{0}^{\phi}\sqrt{1-k^2 \sin^2 \theta}d\theta$ is the incomplete elliptic integral of the second kind, 
due to the non-separability of the equilibrium distribution function, but with a lack of a plasma dispersion function due to the velocity being independent of the absolute value of the momentum as discussed above.
This growth-rate is plotted in the right panel of Fig. \ref{fig:weibel1}.

We can conclude that the behaviour is similar to the already studied case, with a zero real frequency, and a band of unstable modes between $q=0$ and $q_{\rm{crit}}$, with
\begin{equation}
    q_{\rm{crit}} = \frac{e^2 n_e v_F^2}{c^2 \varepsilon T} \frac{\Li{1}(-e^\alpha)}{\Li{2}(-e^\alpha)} \mathrm{E}_0(\gamma) (\gamma^2-1).
\end{equation}
Similar simulations to the quadratic case reveal good agreement when it comes to the overall value of growth-rate of the instability under the given conditions, but with some deviations along the wavenumber axis, as in this case we have only a first order approximation on $\zeta$.

Due to the purely imaginary value of $\omega$, the decay coefficient along $\hat z$ is given by $\kappa_z = \sqrt{q^2+\gamma^2/c^2}$, a real value, therefore showing that, unlike the previously studied waves, these unstable modes are supported and confined in the bidimensional material. 

\subsection{Non-Linear Simulations}

\begin{figure}
    \centering
    \begin{minipage}{0.5\columnwidth}
        \centering
        \includegraphics[width = \textwidth]{"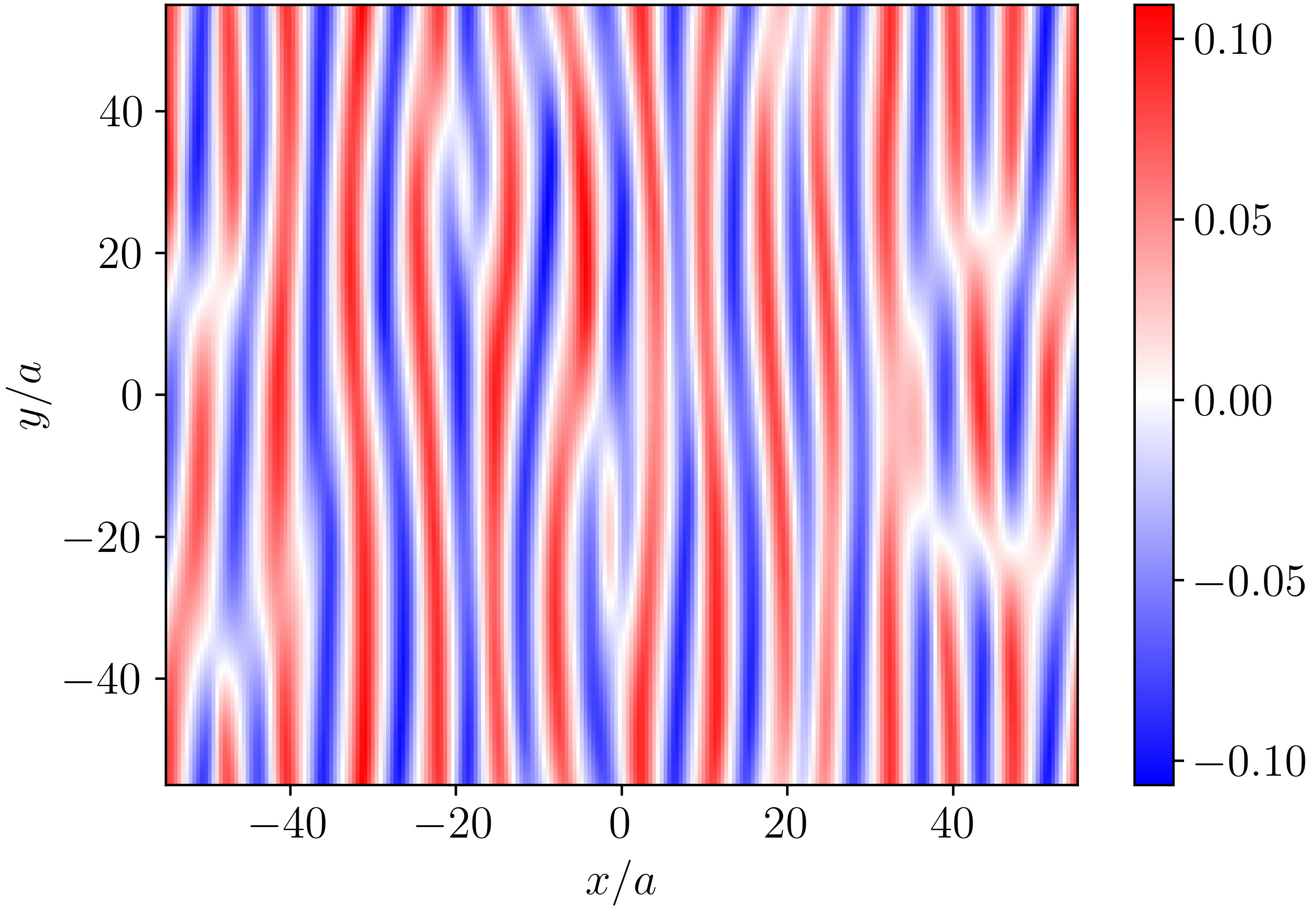"}
    \end{minipage}%
    \begin{minipage}{0.5\columnwidth}
        \centering
        \includegraphics[width = \textwidth]{"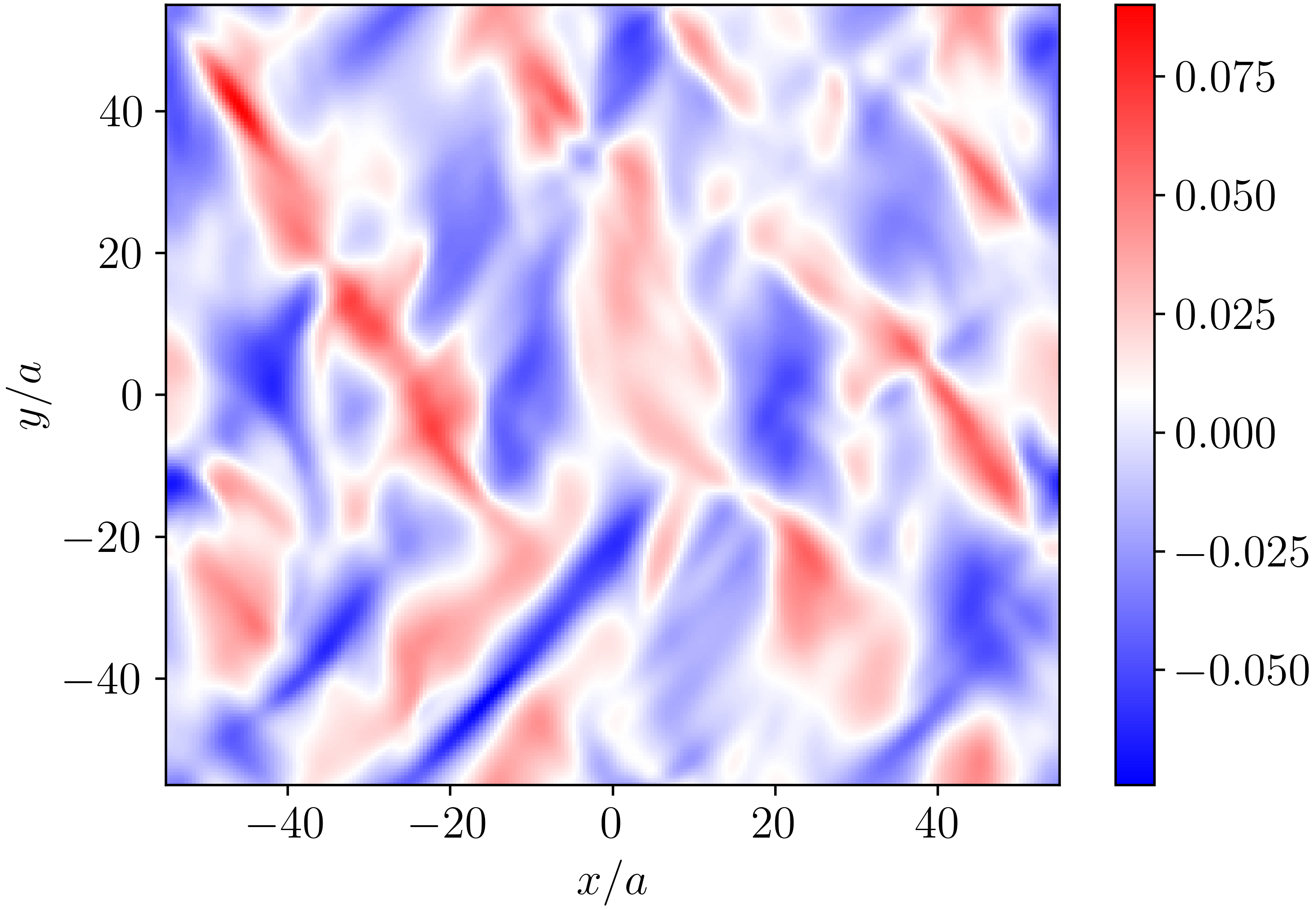"}
    \end{minipage}
    \caption{Transverse (out-of-plane) magnetic field at $t=150 \sqrt{\frac{\varepsilon m}{e^2 n_0^{\frac{3}{2}}}}$ (left) and $t=300 \sqrt{\frac{\varepsilon m}{e^2 n_0^{\frac{3}{2}}}}$ of two-dimensional Weibel instability with quadratic dispersion charge carriers. Temperature of $T = 0.01$, chemical potential $\mu_x = 0.5$ and anisotropy parameter of $\gamma = 5$}
    \label{fig:weibel2Dlin}
\end{figure}

To study the saturation phase of this instability, a number of simulations were performed in four-dimensional phase space, employing both linear and quadratic dispersion relations for the charge carriers, with the results for the out-of-plane ($\hat z$) magnetic field in the linear case being seen in Figure \ref{fig:weibel2Dlin}.

As can be seen, the characteristic magnetic field bands along the direction with smallest chemical potential form as expected. However, in either case, when the non-linear phase is reached, a turbulent breakdown of these structures occurs, giving way to a low amplitude, incoherent magnetic field profile.

This phase is accompanied by a secondary instability, in which charge flows along the wavevector and a spatial dependent electron density forms, in constrast with the transverse nature of the wave in the linear regime. At present then, a method to generate stable magnetic domains has not been found.


\section{Conclusion}
\label{sec:concl}

A kinetic theory of a two-dimensional transverse-electric anisotropy instability  was put forward, in materials with low-energy linear and quadratic bands around charge-neutrality. We demonstrated the existence, growth and confinement of these modes, unlike the known result from TE waves, showing that unlike in the classic three-dimensional case, the growth-rate drops to zero at $q=0$, and deriving the maximum critical wavenumber defining the unstable region.
The theoretical predictions were confirmed numerically, and non-linear simulations show the formation of magnetic domains and their saturation behaviour, involving longitudinal movement of charge carriers and breakdown of the coherent filamentation.
These findings open the possibility of spontaneously generating stable magnetic domains, tunable via properties of the material and imposed externally.


\section*{Acknowledgements}

The authors gratefully acknowledge the computational time allocated on the Deucalion supercomputer by FCT-Fundação para a Ciência e a Tecnologia through project 2025.00002.HPCVLAB.ISTUL.


%
\bibliographystyle{apsrev4-1}

\bibliography{weibel_bibliography}

\begin{thebibliography}{19}%
\makeatletter
\providecommand \@ifxundefined [1]{%
 \@ifx{#1\undefined}
}%
\providecommand \@ifnum [1]{%
 \ifnum #1\expandafter \@firstoftwo
 \else \expandafter \@secondoftwo
 \fi
}%
\providecommand \@ifx [1]{%
 \ifx #1\expandafter \@firstoftwo
 \else \expandafter \@secondoftwo
 \fi
}%
\providecommand \natexlab [1]{#1}%
\providecommand \enquote  [1]{``#1''}%
\providecommand \bibnamefont  [1]{#1}%
\providecommand \bibfnamefont [1]{#1}%
\providecommand \citenamefont [1]{#1}%
\providecommand \href@noop [0]{\@secondoftwo}%
\providecommand \href [0]{\begingroup \@sanitize@url \@href}%
\providecommand \@href[1]{\@@startlink{#1}\@@href}%
\providecommand \@@href[1]{\endgroup#1\@@endlink}%
\providecommand \@sanitize@url [0]{\catcode `\\12\catcode `\$12\catcode
  `\&12\catcode `\#12\catcode `\^12\catcode `\_12\catcode `\%12\relax}%
\providecommand \@@startlink[1]{}%
\providecommand \@@endlink[0]{}%
\providecommand \url  [0]{\begingroup\@sanitize@url \@url }%
\providecommand \@url [1]{\endgroup\@href {#1}{\urlprefix }}%
\providecommand \urlprefix  [0]{URL }%
\providecommand \Eprint [0]{\href }%
\providecommand \doibase [0]{http://dx.doi.org/}%
\providecommand \selectlanguage [0]{\@gobble}%
\providecommand \bibinfo  [0]{\@secondoftwo}%
\providecommand \bibfield  [0]{\@secondoftwo}%
\providecommand \translation [1]{[#1]}%
\providecommand \BibitemOpen [0]{}%
\providecommand \bibitemStop [0]{}%
\providecommand \bibitemNoStop [0]{.\EOS\space}%
\providecommand \EOS [0]{\spacefactor3000\relax}%
\providecommand \BibitemShut  [1]{\csname bibitem#1\endcsname}%
\let\auto@bib@innerbib\@empty
\bibitem [{\citenamefont {Castro~Neto}\ \emph {et~al.}(2009)\citenamefont
  {Castro~Neto}, \citenamefont {Guinea}, \citenamefont {Peres}, \citenamefont
  {Novoselov},\ and\ \citenamefont {Geim}}]{graph_props}%
  \BibitemOpen
  \bibfield  {author} {\bibinfo {author} {\bibfnamefont {A.~H.}\ \bibnamefont
  {Castro~Neto}}, \bibinfo {author} {\bibfnamefont {F.}~\bibnamefont {Guinea}},
  \bibinfo {author} {\bibfnamefont {N.~M.~R.}\ \bibnamefont {Peres}}, \bibinfo
  {author} {\bibfnamefont {K.~S.}\ \bibnamefont {Novoselov}}, \ and\ \bibinfo
  {author} {\bibfnamefont {A.~K.}\ \bibnamefont {Geim}},\ }\href {\doibase
  10.1103/RevModPhys.81.109} {\bibfield  {journal} {\bibinfo  {journal} {Rev.
  Mod. Phys.}\ }\textbf {\bibinfo {volume} {81}},\ \bibinfo {pages} {109}
  (\bibinfo {year} {2009})}\BibitemShut {NoStop}%
\bibitem [{\citenamefont {Aizin}\ and\ \citenamefont {Dyer}(2012)}]{Aizin1}%
  \BibitemOpen
  \bibfield  {author} {\bibinfo {author} {\bibfnamefont {G.~R.}\ \bibnamefont
  {Aizin}}\ and\ \bibinfo {author} {\bibfnamefont {G.~C.}\ \bibnamefont
  {Dyer}},\ }\href {\doibase 10.1103/PhysRevB.86.235316} {\bibfield  {journal}
  {\bibinfo  {journal} {Phys. Rev. B}\ }\textbf {\bibinfo {volume} {86}},\
  \bibinfo {pages} {235316} (\bibinfo {year} {2012})}\BibitemShut {NoStop}%
\bibitem [{\citenamefont {Cosme}\ and\ \citenamefont
  {Ter{\c{c}}as}(2020)}]{cosme_combs}%
  \BibitemOpen
  \bibfield  {author} {\bibinfo {author} {\bibfnamefont {P.}~\bibnamefont
  {Cosme}}\ and\ \bibinfo {author} {\bibfnamefont {H.}~\bibnamefont
  {Ter{\c{c}}as}},\ }\href {\doibase 10.1021/acsphotonics.0c00313} {\bibfield
  {journal} {\bibinfo  {journal} {ACS Photonics}\ }\textbf {\bibinfo {volume}
  {7}},\ \bibinfo {pages} {1375} (\bibinfo {year} {2020})}\BibitemShut
  {NoStop}%
\bibitem [{\citenamefont {Cosme}\ and\ \citenamefont
  {Simões}(2024)}]{cosme_feedback}%
  \BibitemOpen
  \bibfield  {author} {\bibinfo {author} {\bibfnamefont {P.}~\bibnamefont
  {Cosme}}\ and\ \bibinfo {author} {\bibfnamefont {D.}~\bibnamefont
  {Simões}},\ }\href {\doibase 10.1088/1361-648X/ad20a4} {\bibfield  {journal}
  {\bibinfo  {journal} {Journal of Physics: Condensed Matter}\ }\textbf
  {\bibinfo {volume} {36}},\ \bibinfo {pages} {175301} (\bibinfo {year}
  {2024})}\BibitemShut {NoStop}%
\bibitem [{\citenamefont {Stantchev}\ \emph {et~al.}(2017)\citenamefont
  {Stantchev}, \citenamefont {Phillips}, \citenamefont {Hobson}, \citenamefont
  {Hornett}, \citenamefont {Padgett},\ and\ \citenamefont {Hendry}}]{sensing}%
  \BibitemOpen
  \bibfield  {author} {\bibinfo {author} {\bibfnamefont {R.}~\bibnamefont
  {Stantchev}}, \bibinfo {author} {\bibfnamefont {D.}~\bibnamefont {Phillips}},
  \bibinfo {author} {\bibfnamefont {P.}~\bibnamefont {Hobson}}, \bibinfo
  {author} {\bibfnamefont {S.}~\bibnamefont {Hornett}}, \bibinfo {author}
  {\bibfnamefont {M.}~\bibnamefont {Padgett}}, \ and\ \bibinfo {author}
  {\bibfnamefont {E.}~\bibnamefont {Hendry}},\ }\href {\doibase
  10.1364/OPTICA.4.000989} {\bibfield  {journal} {\bibinfo  {journal} {Optica}\
  }\textbf {\bibinfo {volume} {4}},\ \bibinfo {pages} {989} (\bibinfo {year}
  {2017})}\BibitemShut {NoStop}%
\bibitem [{\citenamefont {Fitzgerald}\ \emph {et~al.}(2002)\citenamefont
  {Fitzgerald}, \citenamefont {Berry}, \citenamefont {Zinovev}, \citenamefont
  {Walker}, \citenamefont {Smith},\ and\ \citenamefont
  {Chamberlain}}]{medical}%
  \BibitemOpen
  \bibfield  {author} {\bibinfo {author} {\bibfnamefont {A.~J.}\ \bibnamefont
  {Fitzgerald}}, \bibinfo {author} {\bibfnamefont {E.}~\bibnamefont {Berry}},
  \bibinfo {author} {\bibfnamefont {N.~N.}\ \bibnamefont {Zinovev}}, \bibinfo
  {author} {\bibfnamefont {G.~C.}\ \bibnamefont {Walker}}, \bibinfo {author}
  {\bibfnamefont {M.~A.}\ \bibnamefont {Smith}}, \ and\ \bibinfo {author}
  {\bibfnamefont {J.~M.}\ \bibnamefont {Chamberlain}},\ }\href {\doibase
  10.1088/0031-9155/47/7/201} {\bibfield  {journal} {\bibinfo  {journal}
  {Physics in Medicine and Biology}\ }\textbf {\bibinfo {volume} {47}},\
  \bibinfo {pages} {R67} (\bibinfo {year} {2002})}\BibitemShut {NoStop}%
\bibitem [{\citenamefont {Cosme}\ and\ \citenamefont
  {Ter\ifmmode~\mbox{\c{c}}\else \c{c}\fi{}as}(2023)}]{cosme_nonlinear}%
  \BibitemOpen
  \bibfield  {author} {\bibinfo {author} {\bibfnamefont {P.}~\bibnamefont
  {Cosme}}\ and\ \bibinfo {author} {\bibfnamefont {H.}~\bibnamefont
  {Ter\ifmmode~\mbox{\c{c}}\else \c{c}\fi{}as}},\ }\href {\doibase
  10.1103/PhysRevB.107.195432} {\bibfield  {journal} {\bibinfo  {journal}
  {Phys. Rev. B}\ }\textbf {\bibinfo {volume} {107}},\ \bibinfo {pages}
  {195432} (\bibinfo {year} {2023})}\BibitemShut {NoStop}%
\bibitem [{\citenamefont {Cosme}\ \emph {et~al.}(2023)\citenamefont {Cosme},
  \citenamefont {Terças},\ and\ \citenamefont {Santos}}]{cosme_chiral}%
  \BibitemOpen
  \bibfield  {author} {\bibinfo {author} {\bibfnamefont {P.}~\bibnamefont
  {Cosme}}, \bibinfo {author} {\bibfnamefont {H.}~\bibnamefont {Terças}}, \
  and\ \bibinfo {author} {\bibfnamefont {V.}~\bibnamefont {Santos}},\ }in\
  \href {\doibase 10.1109/NMDC57951.2023.10343623} {\emph {\bibinfo {booktitle}
  {2023 IEEE Nanotechnology Materials and Devices Conference (NMDC)}}}\
  (\bibinfo {year} {2023})\ pp.\ \bibinfo {pages} {908--912}\BibitemShut
  {NoStop}%
\bibitem [{\citenamefont {Nagatsuma}\ \emph {et~al.}(2016)\citenamefont
  {Nagatsuma}, \citenamefont {Ducournau},\ and\ \citenamefont
  {Renaud}}]{comms}%
  \BibitemOpen
  \bibfield  {author} {\bibinfo {author} {\bibfnamefont {T.}~\bibnamefont
  {Nagatsuma}}, \bibinfo {author} {\bibfnamefont {G.}~\bibnamefont
  {Ducournau}}, \ and\ \bibinfo {author} {\bibfnamefont {C.~C.}\ \bibnamefont
  {Renaud}},\ }\href {\doibase 10.1038/nphoton.2016.65} {\bibfield  {journal}
  {\bibinfo  {journal} {Nature Photonics}\ }\textbf {\bibinfo {volume} {10}},\
  \bibinfo {pages} {371} (\bibinfo {year} {2016})}\BibitemShut {NoStop}%
\bibitem [{\citenamefont {Gonçalves}\ and\ \citenamefont
  {Peres}(2016)}]{Peres4}%
  \BibitemOpen
  \bibfield  {author} {\bibinfo {author} {\bibfnamefont {P.~A.}\ \bibnamefont
  {Gonçalves}}\ and\ \bibinfo {author} {\bibfnamefont {N.}~\bibnamefont
  {Peres}},\ }\href {\doibase 10.1142/9948} {\emph {\bibinfo {title} {An
  Introduction to Graphene Plasmonics}}}\ (\bibinfo  {publisher} {World
  Scientific Publishing Company},\ \bibinfo {year} {2016})\ Chap.~\bibinfo
  {chapter} {4}\BibitemShut {NoStop}%
\bibitem [{\citenamefont {Du}\ \emph {et~al.}(2008)\citenamefont {Du},
  \citenamefont {Skachko}, \citenamefont {Barker},\ and\ \citenamefont
  {Andrei}}]{ballistic}%
  \BibitemOpen
  \bibfield  {author} {\bibinfo {author} {\bibfnamefont {X.}~\bibnamefont
  {Du}}, \bibinfo {author} {\bibfnamefont {I.}~\bibnamefont {Skachko}},
  \bibinfo {author} {\bibfnamefont {A.}~\bibnamefont {Barker}}, \ and\ \bibinfo
  {author} {\bibfnamefont {E.~Y.}\ \bibnamefont {Andrei}},\ }\href {\doibase
  10.1038/nnano.2008.199} {\bibfield  {journal} {\bibinfo  {journal} {Nature
  Nanotechnology}\ }\textbf {\bibinfo {volume} {3}},\ \bibinfo {pages} {491}
  (\bibinfo {year} {2008})}\BibitemShut {NoStop}%
\bibitem [{\citenamefont {Narozhny}\ \emph {et~al.}(2017)\citenamefont
  {Narozhny}, \citenamefont {Gornyi}, \citenamefont {Mirlin},\ and\
  \citenamefont {Schmalian}}]{hydrodynamic}%
  \BibitemOpen
  \bibfield  {author} {\bibinfo {author} {\bibfnamefont {B.~N.}\ \bibnamefont
  {Narozhny}}, \bibinfo {author} {\bibfnamefont {I.~V.}\ \bibnamefont
  {Gornyi}}, \bibinfo {author} {\bibfnamefont {A.~D.}\ \bibnamefont {Mirlin}},
  \ and\ \bibinfo {author} {\bibfnamefont {J.}~\bibnamefont {Schmalian}},\
  }\href {\doibase https://doi.org/10.1002/andp.201700043} {\bibfield
  {journal} {\bibinfo  {journal} {Annalen der Physik}\ }\textbf {\bibinfo
  {volume} {529}},\ \bibinfo {pages} {1700043} (\bibinfo {year}
  {2017})}\BibitemShut {NoStop}%
\bibitem [{\citenamefont {Moyal}(1949)}]{moyal1949}%
  \BibitemOpen
  \bibfield  {author} {\bibinfo {author} {\bibfnamefont {J.~E.}\ \bibnamefont
  {Moyal}},\ }\href {\doibase 10.1017/S0305004100000487} {\bibfield  {journal}
  {\bibinfo  {journal} {Mathematical Proceedings of the Cambridge Philosophical
  Society}\ }\textbf {\bibinfo {volume} {45}},\ \bibinfo {pages} {99} (\bibinfo
  {year} {1949})}\BibitemShut {NoStop}%
\bibitem [{\citenamefont {Weibel}(1959)}]{Weibel}%
  \BibitemOpen
  \bibfield  {author} {\bibinfo {author} {\bibfnamefont {E.~S.}\ \bibnamefont
  {Weibel}},\ }\href {\doibase 10.1103/PhysRevLett.2.83} {\bibfield  {journal}
  {\bibinfo  {journal} {Phys. Rev. Lett.}\ }\textbf {\bibinfo {volume} {2}},\
  \bibinfo {pages} {83} (\bibinfo {year} {1959})}\BibitemShut {NoStop}%
\bibitem [{\citenamefont {Fried}(1959)}]{Fried}%
  \BibitemOpen
  \bibfield  {author} {\bibinfo {author} {\bibfnamefont {B.~D.}\ \bibnamefont
  {Fried}},\ }\href {\doibase 10.1063/1.1705933} {\bibfield  {journal}
  {\bibinfo  {journal} {The Physics of Fluids}\ }\textbf {\bibinfo {volume}
  {2}},\ \bibinfo {pages} {337} (\bibinfo {year} {1959})}\BibitemShut {NoStop}%
\bibitem [{\citenamefont {Qi}\ and\ \citenamefont {Lucas}(2021)}]{trilayer}%
  \BibitemOpen
  \bibfield  {author} {\bibinfo {author} {\bibfnamefont {M.}~\bibnamefont
  {Qi}}\ and\ \bibinfo {author} {\bibfnamefont {A.}~\bibnamefont {Lucas}},\
  }\href {\doibase 10.1103/PhysRevB.104.195106} {\bibfield  {journal} {\bibinfo
   {journal} {Phys. Rev. B}\ }\textbf {\bibinfo {volume} {104}},\ \bibinfo
  {pages} {195106} (\bibinfo {year} {2021})}\BibitemShut {NoStop}%
\bibitem [{\citenamefont {Fu}\ \emph {et~al.}(2018)\citenamefont {Fu},
  \citenamefont {Scaffidi}, \citenamefont {Waissman}, \citenamefont {Sun},
  \citenamefont {Saha}, \citenamefont {Watzman}, \citenamefont {Srivastava},
  \citenamefont {Li}, \citenamefont {Schnelle}, \citenamefont {Werner},
  \citenamefont {Kamminga}, \citenamefont {Sachdev}, \citenamefont {Parkin},
  \citenamefont {Hartnoll}, \citenamefont {Felser},\ and\ \citenamefont
  {Gooth}}]{anisotropic1}%
  \BibitemOpen
  \bibfield  {author} {\bibinfo {author} {\bibfnamefont {C.}~\bibnamefont
  {Fu}}, \bibinfo {author} {\bibfnamefont {T.}~\bibnamefont {Scaffidi}},
  \bibinfo {author} {\bibfnamefont {J.}~\bibnamefont {Waissman}}, \bibinfo
  {author} {\bibfnamefont {Y.}~\bibnamefont {Sun}}, \bibinfo {author}
  {\bibfnamefont {R.}~\bibnamefont {Saha}}, \bibinfo {author} {\bibfnamefont
  {S.~J.}\ \bibnamefont {Watzman}}, \bibinfo {author} {\bibfnamefont {A.~K.}\
  \bibnamefont {Srivastava}}, \bibinfo {author} {\bibfnamefont
  {G.}~\bibnamefont {Li}}, \bibinfo {author} {\bibfnamefont {W.}~\bibnamefont
  {Schnelle}}, \bibinfo {author} {\bibfnamefont {P.}~\bibnamefont {Werner}},
  \bibinfo {author} {\bibfnamefont {M.~E.}\ \bibnamefont {Kamminga}}, \bibinfo
  {author} {\bibfnamefont {S.}~\bibnamefont {Sachdev}}, \bibinfo {author}
  {\bibfnamefont {S.~S.~P.}\ \bibnamefont {Parkin}}, \bibinfo {author}
  {\bibfnamefont {S.~A.}\ \bibnamefont {Hartnoll}}, \bibinfo {author}
  {\bibfnamefont {C.}~\bibnamefont {Felser}}, \ and\ \bibinfo {author}
  {\bibfnamefont {J.}~\bibnamefont {Gooth}},\ }\href@noop {} {\enquote
  {\bibinfo {title} {Thermoelectric signatures of the electron-phonon fluid in
  {{PtSn\textsubscript{4}}}},}\ } (\bibinfo {year} {2018}),\ \Eprint
  {http://arxiv.org/abs/1802.09468} {arXiv:1802.09468 [cond-mat.mtrl-sci]}
  \BibitemShut {NoStop}%
\bibitem [{\citenamefont {Vool}\ \emph {et~al.}(2021)\citenamefont {Vool},
  \citenamefont {Hamo}, \citenamefont {Varnavides}, \citenamefont {Wang},
  \citenamefont {Zhou}, \citenamefont {Kumar}, \citenamefont {Dovzhenko},
  \citenamefont {Qiu}, \citenamefont {Garcia}, \citenamefont {Pierce},
  \citenamefont {Gooth}, \citenamefont {Anikeeva}, \citenamefont {Felser},
  \citenamefont {Narang},\ and\ \citenamefont {Yacoby}}]{anisotropic2}%
  \BibitemOpen
  \bibfield  {author} {\bibinfo {author} {\bibfnamefont {U.}~\bibnamefont
  {Vool}}, \bibinfo {author} {\bibfnamefont {A.}~\bibnamefont {Hamo}}, \bibinfo
  {author} {\bibfnamefont {G.}~\bibnamefont {Varnavides}}, \bibinfo {author}
  {\bibfnamefont {Y.}~\bibnamefont {Wang}}, \bibinfo {author} {\bibfnamefont
  {T.~X.}\ \bibnamefont {Zhou}}, \bibinfo {author} {\bibfnamefont
  {N.}~\bibnamefont {Kumar}}, \bibinfo {author} {\bibfnamefont
  {Y.}~\bibnamefont {Dovzhenko}}, \bibinfo {author} {\bibfnamefont
  {Z.}~\bibnamefont {Qiu}}, \bibinfo {author} {\bibfnamefont {C.~A.~C.}\
  \bibnamefont {Garcia}}, \bibinfo {author} {\bibfnamefont {A.~T.}\
  \bibnamefont {Pierce}}, \bibinfo {author} {\bibfnamefont {J.}~\bibnamefont
  {Gooth}}, \bibinfo {author} {\bibfnamefont {P.}~\bibnamefont {Anikeeva}},
  \bibinfo {author} {\bibfnamefont {C.}~\bibnamefont {Felser}}, \bibinfo
  {author} {\bibfnamefont {P.}~\bibnamefont {Narang}}, \ and\ \bibinfo {author}
  {\bibfnamefont {A.}~\bibnamefont {Yacoby}},\ }\href {\doibase
  10.1038/s41567-021-01341-w} {\bibfield  {journal} {\bibinfo  {journal}
  {Nature Physics}\ }\textbf {\bibinfo {volume} {17}},\ \bibinfo {pages} {1216}
  (\bibinfo {year} {2021})}\BibitemShut {NoStop}%
\bibitem [{\citenamefont {Moll}\ \emph {et~al.}(2016)\citenamefont {Moll},
  \citenamefont {Kushwaha}, \citenamefont {Nandi}, \citenamefont {Schmidt},\
  and\ \citenamefont {Mackenzie}}]{anisotropic3}%
  \BibitemOpen
  \bibfield  {author} {\bibinfo {author} {\bibfnamefont {P.~J.~W.}\
  \bibnamefont {Moll}}, \bibinfo {author} {\bibfnamefont {P.}~\bibnamefont
  {Kushwaha}}, \bibinfo {author} {\bibfnamefont {N.}~\bibnamefont {Nandi}},
  \bibinfo {author} {\bibfnamefont {B.}~\bibnamefont {Schmidt}}, \ and\
  \bibinfo {author} {\bibfnamefont {A.~P.}\ \bibnamefont {Mackenzie}},\ }\href
  {\doibase 10.1126/science.aac8385} {\bibfield  {journal} {\bibinfo  {journal}
  {Science}\ }\textbf {\bibinfo {volume} {351}},\ \bibinfo {pages} {1061}
  (\bibinfo {year} {2016})}\BibitemShut {NoStop}%
\end{thebibliography}%

\end{document}